\documentclass[11pt]{article}
\usepackage{aas_macros,amsmath,amssymb,comment,cite,esint,graphicx,mathtools,diagbox}
\usepackage{bm}
\usepackage[margin=.8in,letterpaper]{geometry}
\usepackage[colorlinks=true]{hyperref}
\usepackage[affil-it]{authblk}
\usepackage{subcaption}
\usepackage[utf8]{inputenc}
\usepackage{mathrsfs}
\usepackage{appendix}
\usepackage{amssymb}
\usepackage{float}                  
\usepackage{color}
\usepackage{cite}
\usepackage{hyperref}
\hypersetup{pageanchor=false}
\usepackage{indentfirst}
\usepackage{url}
\usepackage{xfrac}
\usepackage{caption}
\usepackage[numbers,square,comma,sort&compress,merge]{natbib}
\usepackage{esint}
\usepackage{overpic}
\usepackage{graphicx}
\usepackage{epsf,amsmath,bbold,amsfonts,stmaryrd}
\usepackage{textcomp}
\usepackage{ulem}
\usepackage{tikz}
\usepackage{multirow}
\numberwithin{equation}{section}
\let\originalleft\left
\let\originalright\right
\renewcommand{\left}{\mathopen{}\mathclose\bgroup\originalleft}
\renewcommand{\right}{\aftergroup\egroup\originalright}

\def\bea{\begin{eqnarray}}
\def\eea{\end{eqnarray}}
\def\nn{\nonumber}

\usepackage{tensor}
\usepackage{physics}

\newcolumntype{P}[1]{>{\Centering\hspace{0pt}}p{#1}}
\newcolumntype{Z}{>{\centering\arraybackslash}X} 

\newcommand{\codename}{\textbf{Coport}}

\newcommand{\br}[1]{\left[#1\right]}
\newcommand{\ee}{\mathrm{e}}
\newcommand{\Mc}[1]{\mathcal{#1}}
\newcommand{\mJ}{\mathcal{J}}
\newcommand{\mA}{\mathcal{A}}
\newcommand{\mR}{\mathcal{R}}

\newcommand{\mD}{\mathcal{D}}
\newcommand{\mX}{\mathcal{X}}
\newcommand{\mP}{\mathcal{P}}
\newcommand{\mQ}{\mathcal{Q}}
\newcommand{\mE}{\mathcal{E}}
\newcommand{\mY}{\mathcal{Y}}

\newcommand{\dif}{\mathrm{d}}   
   
\newcommand{\lrg}[1]{\langle #1\rangle }
\newcommand{\pa}[1]{\left(#1\right)}

\begin{document}
\title{\codename: A New Public Code for Polarized Radiative Transfer in a Covariant Framework$^\spadesuit$}

\author{
Jiewei Huang$^{1}$, Liheng Zheng$^{1}$, Minyong Guo$^{1,2\ast}$, Bin Chen$^{3,4,5}$}
\date{}

\maketitle

\vspace{-10mm}

\begin{center}
{\it
$^1$ School of physics and astronomy, Beijing Normal University,
Beijing 100875, P. R. China\\\vspace{4mm}

$^2$Key Laboratory of Multiscale Spin Physics (Ministry of Education), Beijing Normal University, Beijing 100875, China\\\vspace{4mm}

$^3$ Department of Physics, School of Physical Science and Technology, Ningbo University, Ningbo, Zhejiang 315211, China\\\vspace{4mm}

$^4$Department of Physics, Peking University, No.5 Yiheyuan Rd, Beijing
100871, P.R. China\\\vspace{4mm}

$^5$Center for High Energy Physics, Peking University,
No.5 Yiheyuan Rd, Beijing 100871, P. R. China\\\vspace{2mm}

\vspace{2mm}
}
\end{center}

\vspace{8mm}

\begin{abstract}

General relativistic radiative transfer calculations are essential for comparing theoretical models of black hole accretion flows and jets with observational data. In this work, we introduce \codename, a novel public code specifically designed for covariant polarized ray-tracing radiative transfer computations in any spacetime. Written in Julia, \codename\ includes an interface for visualizing numerical results obtained from HARM, a publicly available implementation of the general relativistic magnetohydrodynamics code. We validate the precision of our code by comparing its outputs with the results from a variety of established methodologies. This includes the verification against analytical solutions, the validation through thin-disk assessments, and the evaluation via thick-disk analyses. Notably, our code employs a methodology that eliminates the need for separating the computations of spacetime propagation and plasma propagation. Instead, it directly solves the coupled, covariant, polarized radiative transfer equation in curved spacetime, seamlessly integrating the effects of gravity with plasma influences. This approach sets our code apart from the existing alternatives and enhances its accuracy and efficiency.

\end{abstract}

\vfill{\footnotesize $\ast$ Corresponding author: minyongguo@bnu.edu.cn; \par$^\spadesuit$ The public version of \codename\ is available at the following URL: https://github.com/JieweiHuang/Coport.}

\maketitle

\newpage
\baselineskip 18pt

\section{Introduction}\label{sec1}

The captured images of black holes using the Event Horizon Telescope (EHT) not only reveals the intensity distribution but also provides crucial details about the polarization of light rays \cite{EventHorizonTelescope:2019dse, EventHorizonTelescope:2022wkp, EventHorizonTelescope:2021bee, EventHorizonTelescope:2019pgp, EventHorizonTelescope:2021srq}. To  understand the polarized images of black holes theoretically, it is essential to use general relativistic radiative transfer (GRRT) techniques for imaging their accreting plasma. This makes GRRT an indispensable cornerstone technology in the study of black hole images.

When calculating the total intensity of a light source, the process within GRRT is relatively straightforward. It involves solving the geodesic equations for light rays and an additional radiative transfer equation to determine the intensity. Relevant techniques can be found in \cite{Gold:2020iql} and in our previous work \cite{Hu:2020usx, Hou:2022eev, Zhang:2024lsf}. However, the complexity in crafting the polarization images of a black hole increases to another level. Even though the trajectories of light rays still follow the geodesics and the polarization vectors undergo parallel transport along null geodesics in curved spacetime,  the radiative transfer equation extends beyond the simple intensity formulation,  and involves linear and circular polarization parameters along with the intensity. This significantly escalates the complexity of the computations.

In the field of ray-tracing code for covariant, polarized radiative transport, there are several frameworks currently in use \cite{EventHorizonTelescope:2023hqy}, including grtrans \cite{Dexter:2009fg, Dexter:2016cdk}, ipole \cite{Pfrommer:2021uzq, Moscibrodzka:2017lcu}, RAPTOR \cite{Bronzwaer:2018lde, Bronzwaer:2020kle}, Odyssey \cite{Pu:2016eml, Pu:2018ute}, and BHOSS \cite{Younsi:2012ha, Younsi:2019iee}, among others \cite{Aimar:2023vcs, pelle2022skylight}. A common feature among these computational methods is the separation of equations governing the gravitational influence on light rays from those describing the interaction of light rays with the plasma. Initially, the parallel transport of the polarization vector along null geodesics in curved spacetime are solved, disregarding plasma effects. The updated polarization vector is then used as input for the next stage. This process continues by applying the evolution equations for the Stokes parameters in flat spacetime within the fluid regime, ultimately leading to the final outcomes. Nevertheless, there are distinctions in the specific handling of details. For instance, in the case of grtrans, the parallel transport of the polarization vector relies on the Penrose-Walker constant, which exists in Type D spacetimes, making it unsuitable for arbitrary spacetimes. In contrast, BHOSS initially represents the observer's polarization basis using parallel and perpendicular 4-vectors before proceeding with the computations. RAPTOR, on the other hand, has devised a concise, Lorentz-invariant representation of a polarized ray. Similar to BHOSS, Odyssey operates as a GPU-based code. A distinctive approach is seen in ipole, which originates from a covariant equation that couples both gravitational and plasma influences \cite{2012ApJ...752..123G}. However, in its implementation, ipole separates the coupled equations into gravitational and plasma components, thus adopting a two-step process within sufficiently small steps. 

In light of the covariant equation that intertwines gravitational and plasma influences \cite{2012ApJ...752..123G}, we aim to pioneer a novel approach and develop a code that achieves a unified evolution of the coupled equations. Different from ipole, we do not try to decouple the covariant equation into two components. Our endeavor leads to the code \codename. We will  demonstrate how it accomplishes a one-step evolution of both spacetime and plasma propagation, and we make  a thorough examination of the precision and accuracy of \codename. Although our computations use the Kerr black hole as an example, it is important to note that \codename\ can be applied to any spacetime.

The remaining parts of the paper are organized as follows. In Sec. \ref{sec2}, we elucidate the polarized radiation transfer equations and their specific treatment methodologies. In Sec. \ref{sec3} we discuss various accretion disk models, detailing both the numerical schemes and code verification processes. In Sec. \ref{sec4} we provide a summary and discuss  our research findings. In this work, we will use the units where \( G = c = 1 \), unless otherwise specified.

\section{Polarized radiative transfer in curved spacetime}\label{sec2}

In the Boyer-Lindquist (BL) coordinates, the metric for Kerr spacetime is given by
\begin{align}
    \dif s^2 = -\dfrac{\Delta}{\Sigma}
    \left(\dif t - a\sin^2\theta \dif\phi\right)^2
    + \Sigma \left(\dfrac{\dif r^2}{\Delta} + \dif \theta^2\right)
    + \dfrac{\sin^2\theta}{\Sigma}
    \left[a \dif t - (r^2 + a^2) \dif\phi\right]^2\,,
	\label{eqn:downmatrices}
\end{align}
where  
\begin{align}
\Delta = r^2 - 2Mr + a^2\,, \quad \quad \Sigma = r^2 + a^2 \cos^2\theta\,,
\end{align}
with $M$ and $a$ being the mass parameter and the spin parameter, respectively. The existence of inner and outer horizons in a Kerr black hole can be determined by solving the equation \(\Delta(r_{\pm}) = 0\). The outer horizon, also known as the event horizon, satisfies
\begin{align}
	r_{+}=1+\sqrt{1-a^2}\,.
\end{align}
For convenience, we set \( M = 1 \) from this point forward. 

To image the luminous matter surrounding a black hole, it is essential to understand both the trajectory of light rays and the radiative transfer of light. We numerically solve the Hamiltonian canonical form of the geodesic equation using a backward ray-tracing method to determine the paths of light rays. Imaging is achieved through fisheye camera projection. For specific technical details, please refer to Appendix B in \cite{Hu:2020usx}. To describe the interaction between light rays and matter in radiative transfer, this study employs the tensor form of the radiative transfer equation as presented in \cite{2012ApJ...752..123G}:
\begin{align}
    k^\mu\nabla_\mu \mathcal{S}^{\alpha\beta} = \mathcal{J}^{\alpha\beta} + H^{\alpha\beta\mu\nu}\mathcal{S}_{\mu\nu}\,.
    \label{eqn:eqn1}
\end{align}
Here, \(k^\mu\) is the wave vector for photons, \(\mathcal{J}^{\alpha\beta}\) characterizes the emission from the source, and \(H^{\alpha\beta\mu\nu}\) represents absorption and Faraday rotation effects. The quantity \(\mathcal{S}^{\alpha\beta}\), referred to as the polarization tensor, describes polarization as a Hermitian tensor, satisfying \(\mathcal{S}^{\alpha\beta} = \bar{\mathcal{S}}^{\beta\alpha}\), where the bar denotes complex conjugation. The details on the derivation of Eq. (\ref{eqn:eqn1}), as well as the definitions of \(H^{\alpha\beta\mu\nu}\) and the properties of \(\mathcal{S}^{\alpha\beta}\), can be found in Appendix \ref{sec:A}. This allows us to avoid further elaboration in the main text.

In practical computations, particularly when analyzing accretion disk material near black holes, it is crucial to establish a specific frame of reference to accurately depict the emission, the absorption, and the rotation of local polarized light. A suitable fluid coordinate system, as delineated in \cite{Broderick:2003fc}, is adopted in this work. Given a fluid four-velocity \( u^\mu \), a light ray wavenumber \( k^\mu \), and any spacelike vector \( d^\mu \), the four basis vectors of this coordinate system are respectively
\begin{align}
    e^{\mu}_{(0)}=u^\mu\,,\quad e^{\mu}_{(3)}=\dfrac{k^\mu}{\omega}-u^\mu\,,\quad e^{\mu}_{(2)}=\dfrac{1}{\mathcal{N}}\left(
        d^\mu+\beta u^\mu-C e^\mu_{(3)}
    \right)\,,\quad e^{\mu}_{(1)}=\dfrac{\epsilon^{\mu\nu\sigma\rho}u_\nu k_\sigma d_\rho }{\omega \mathcal{N}}\,,\label{cflu}
\end{align}
where, $\epsilon^{\mu\nu\sigma\rho}$ is the Levi-Civita tensor, with 
\begin{align}
    \begin{array}{lllll}
        d^2=d_\mu d^\mu \,,
        &\beta=u_\mu d^\mu \,,
        &\omega=-k_\mu u^\mu\,, 
        &C=\dfrac{k_\mu d^\mu}{\omega}-\beta\,,
        &\mathcal{N}=\sqrt{d^2+\beta^2-C^2}\,.
    \end{array}
\end{align}
We typically set \( d^\mu \) as the local magnetic field \( b^\mu \). This choice leads to the vanishing of all emission, absorption, and rotation coefficients associated with the Stokes parameter $U$ in the context of polarization. Below, we will explore the representations of various tensors within the fluid coordinate system. 

Taking into account the gauge symmetry inherent in the polarization tensor \(\mathcal{S}^{\alpha\beta}\), we can restrict our discussion of the tensors to the orthogonal subspace \(\{e_{(1)}^\mu, e_{(2)}^\mu\}\). The projection of \(\mathcal{S}^{\alpha\beta}\) within the orthogonal subspace \(\{e_{(1)}^\mu, e_{(2)}^\mu\}\) is given by
\bea
\hat{S}^{(a)(b)}=
\begin{pmatrix}
\mathcal{I} + \mathcal{Q} & \mathcal{U} + i\mathcal{V} \\
\mathcal{U} - i\mathcal{V} & \mathcal{I} - \mathcal{Q}
\end{pmatrix}.\label{Spro}
\eea
Here, \(\{\mathcal{I}, \mathcal{Q}, \mathcal{U}, \mathcal{V}\} = \left\{ I/\nu^3, Q/\nu^3, U/\nu^3, V/\nu^3 \right\}\) represent the invariant Stokes parameters as shown in Eq. (\ref{subs}). The hat notation indicates the projection in the orthogonal subspace \(\{e_{(1)}^\mu, e_{(2)}^\mu\}\). In the local frame,  the local Stokes parameters follow the evolution equation given below
\begin{align}
\dfrac{\dif}{\dif \lambda}
    \begin{pmatrix}
        \Mc{I} \\ \Mc{Q}\\ \Mc{U}\\ \Mc{V}
    \end{pmatrix}
    =\dfrac{1}{\omega^2}\begin{pmatrix}
        j_{I} \\ j_{Q}\\ j_{U}\\ j_{V}
    \end{pmatrix}
    -{\omega}\begin{pmatrix}
        a_I &a_Q &a_U &a_V\\
        a_Q &a_I &r_V &-r_U\\
        a_U &-r_V &a_I &r_Q\\
        a_V &r_U &-r_Q &a_I
    \end{pmatrix}
    \begin{pmatrix}
        \Mc{I} \\ \Mc{Q}\\ \Mc{U}\\ \Mc{V}
    \end{pmatrix}\,,
    \label{eqn:stksydfcwp}
\end{align}
where, $\lambda$ denotes the affine parameter of photons, and \(\omega = -k_\mu u^\mu\) represents the frequency of photons as observed by a co-moving observer within the fluid. The emission, absorption, and Faraday rotation coefficients $j$, $a$ and $r$ appearing in the Eq. (\ref{eqn:stksydfcwp}) are described in detail in Appendix \ref{sec:FC}. By comparing the projection of Eq. (\ref{eqn:eqn1}) in the rest frame of the fluid with Eq. (\ref{eqn:stksydfcwp}), we can derive the projection of the emission tensor $\mJ^{(a)(b)}$ in the orthogonal subspace of the fluid frame, denoted by \(\{e_{(1)}^\mu, e_{(2)}^\mu\}\), as follows\cite{1996A&AS..117..161H}:
\begin{align}
\hat{\mJ}^{(a)(b)}=\dfrac{1}{\omega^2}
    \begin{pmatrix}
        j_{I}+j_{Q} &j_{U}+ij_{V}\\
        j_{U}-ij_{V}& j_{I}-j_{Q}
      \end{pmatrix}\,.
      \label{eqn:fsxs}
\end{align}
Furthermore, considering Eq. (\ref{hdec}), the term \( H^{\alpha\beta\mu\nu} \) can be decomposed into two components: one representing the absorption, denoted as \( \mA^{\alpha\beta\mu\nu} \), and the other representing the Faraday rotation, denoted as \( \mR^{\alpha\beta\mu\nu} \). By comparing this with Eq. (\ref{eqn:stksydfcwp}), we can similarly derive the absorption and rotation coefficients within a local orthogonal subspace as follows

\begin{align}
    \hat{\mA}^{(a)(b)} = -\omega
    \begin{pmatrix}
        a_{I} + a_{Q} & a_{U} + ia_{V} \\
        a_{U} - ia_{V} & a_{I} - a_{Q}
    \end{pmatrix}
\,,\qquad
    \hat{\mR}^{(a)(b)} = i\omega
    \begin{pmatrix}
        r_{Q} & r_{U} + ir_{V} \\
        r_{U} - ir_{V} & -r_{Q}
    \end{pmatrix}
    \,.\label{eqn:xzxx}
\end{align}
It is worth noting that, due to the presence of gauge symmetries, we can set all coefficients outside the orthogonal subspace in $\mJ^{(a)(b)}, \mA^{(a)(b)}$ and $\mR^{(a)(b)}$ to zero. At this point, all tensors in Eq. (\ref{eqn:eqn1})—\(\mathcal{S}^{\alpha\beta}\), \(\mathcal{J}^{\alpha\beta}\), and \(H^{\alpha\beta\mu\nu}\)—have been determined. 

Next, we proceed to solve Eq. (\ref{eqn:eqn1}). First, we reformulate Equation (\ref{eqn:eqn1}) into a first-order differential equation
\begin{align}
   \dot{\Mc{S}}^{\alpha\beta}=
    -\tensor{\Gamma}{^\alpha_\mu_\nu}k^\mu \Mc{S}^{\nu\beta}
    -\tensor{\Gamma}{^\beta_\mu_\nu}k^\mu \Mc{S}^{\alpha\nu}+
    \Mc{J}^{\alpha\beta}+H^{\alpha\beta\mu\nu}\Mc{S}_{\mu\nu}\,,
    \label{eqn:eqn1fenjiexingshi}
\end{align}
where, the dot denotes the differentiation with respect to the photon's affine parameter, \(\dfrac{\dif}{\dif\lambda}\), while \(\tensor{\Gamma}{^\alpha_\mu_\nu}\) represents the Christoffel symbols in the corresponding coordinate system. By definition, \(\mathcal{S}^{\alpha\beta}\), \(\mathcal{J}^{\alpha\beta}\), and \(H^{\alpha\beta\mu\nu}\) are all complex tensors. To facilitate numerical computations, our approach involves first transforming Equation (\ref{eqn:eqn1}) into a real-valued form. 

Considering the fact that \(\mathcal{S}^{\alpha\beta} = \bar{\mathcal{S}}^{\beta\alpha}\) is Hermitian, the tensor \(\mathcal{S}^{\alpha\beta}\) possesses only 16 degrees of freedom. It can be decomposed into two components as follows:
\begin{align}
\mathcal{S}^{\alpha\beta} = \mathcal{D}^{\alpha\beta} + i\mathcal{X}^{\alpha\beta}\,,\label{Scom}
\end{align}
Here, \(\mathcal{D}^{\alpha\beta} = \mathcal{D}^{\beta\alpha}\) describes the total intensity and linearly polarized light, with 10 degrees of freedom. On the other hand, \(\mathcal{X}^{\alpha\beta} = -\mathcal{X}^{\beta\alpha}\) characterizes circularly polarized light, containing 6 degrees of freedom.
By taking the real and imaginary parts of Eq. (\ref{eqn:eqn1fenjiexingshi}), we can split it into two sets of equations:
\begin{align}
    \dot{\mD}^{\alpha\beta}&=
    -\tensor{\Gamma}{^\alpha_\mu_\nu}k^\mu \mD^{\nu\beta}
    -\tensor{\Gamma}{^\beta_\mu_\nu}k^\mu \mD^{\alpha\nu}+
    \mE^{\alpha\beta} +\dfrac{1}{2}\pa{
        \mP^{\alpha\mu}\tensor{\mD}{_\mu^\beta}+
        \mD^{\alpha\mu}\tensor{\mP}{^\beta_\mu}-
        \mQ^{\alpha\mu}\tensor{\mX}{_\mu^\beta}+
        \mX^{\alpha\mu}\tensor{\mQ}{^\beta_\mu}
    }
    \,,\label{eqn:fszyfcnew0}
\\
    \dot\mX^{\alpha\beta}&=
    -\tensor{\Gamma}{^\alpha_\mu_\nu}k^\mu \mX^{\nu\beta}
    -\tensor{\Gamma}{^\beta_\mu_\nu}k^\mu \mX^{\alpha\nu}+
    \mY^{\alpha\beta} +\dfrac{1}{2}\pa{
        \mP^{\alpha\mu}\tensor{\mX}{_\mu^\beta}+
        \mX^{\alpha\mu}\tensor{\mP}{^\beta_\mu}+
        \mQ^{\alpha\mu}\tensor{\mD}{_\mu^\beta}-
        \mD^{\alpha\mu}\tensor{\mQ}{^\beta_\mu}
    }\,.\label{eqn:fszyfcnew}
\end{align}
Here, for notational convenience, we have introduced four real tensors \( \mE^{\alpha\beta} \), \( \mY^{\alpha\beta} \), \( \mP^{\alpha\beta} \), and \( \mQ^{\alpha\beta} \), which satisfy the following conditions
\begin{align}
    \mE^{\alpha\beta}=\dfrac{1}{2}\pa{\mJ^{\alpha\beta}+\bar{\mJ}^{\alpha\beta}}\,,\quad  \mY^{\alpha\beta}=\dfrac{1}{2i}\pa{\mJ^{\alpha\beta}-\bar{\mJ}^{\alpha\beta}}\,,
    \label{eqn:Jjbs}
\end{align}
and 
\begin{align}
    \mP^{\alpha\beta}=\dfrac{1}{2}\pa{\mA^{\alpha\beta}+\bar{\mA}^{\alpha\beta}+\mR^{\alpha\beta}+\bar \mR^{\alpha\beta}}\,,\quad \mQ^{\alpha\beta}=\dfrac{1}{2i}\pa{\mA^{\alpha\beta}-\bar{\mA}^{\alpha\beta}+\mR^{\alpha\beta}-\bar{\mR}^{\alpha\beta}}\,.\label{mPQ}
\end{align}
Clearly, we observe that \(\mE^{\alpha\beta}\) and \(\mY^{\alpha\beta}\) represent the real and imaginary parts, respectively, of \(\mJ^{\alpha\beta}\). Similarly, \(\mP^{\alpha\beta}\) and \(\mQ^{\alpha\beta}\) are the sums of the real and imaginary parts, respectively, of \(\mA^{\alpha\beta}\) and \(\mR^{\alpha\beta}\).

\begin{figure}[ht]
    \centering
    \includegraphics[width=5.5in]{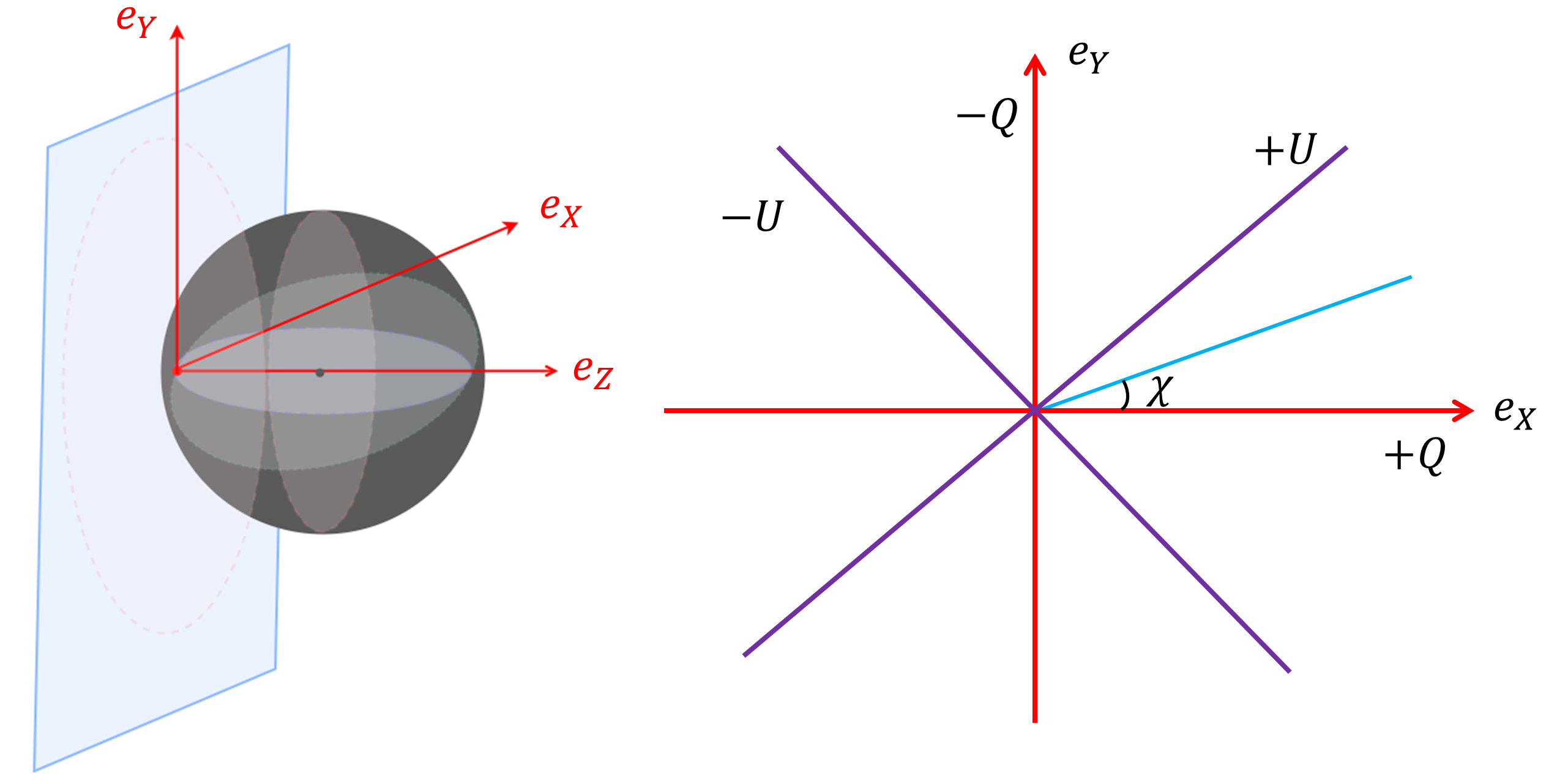}
    \caption{Illustration of the screen and Stokes parameters. Here, we adhere to the `looking outward' gauge as per the IAU standards, where \( e_Z \propto \partial_r \) points outward, while \( e_X \) and \( e_Y \) form the basis of the screen.}
    \label{fig:IAUconvent}
\end{figure}

With the aid of Eqs. (\ref{Spro}), (\ref{eqn:fsxs}), (\ref{eqn:xzxx}), (\ref{Scom}), (\ref{eqn:Jjbs}), and (\ref{mPQ}), all quantities in Eqs. (\ref{eqn:fszyfcnew0}) and (\ref{eqn:fszyfcnew}) have now been explicitly and concretely expressed. Consequently, we are now prepared to proceed with the resolution. Solving Eqs. (\ref{eqn:fszyfcnew0}) and (\ref{eqn:fszyfcnew}) requires the use of a coordinate system. For non-local equations evolving in spacetime, the BL coordinate system is particularly suitable. However, since the emission coefficient \(\mJ^{\alpha\beta}\), the absorption coefficient \(\mA^{\alpha\beta}\), and the rotation coefficients \(\mR^{\alpha\beta}\) are initially provided in a local fluid system, it is essential to transform these coefficients into the ones in the BL coordinate system first. The procedure for this transformation involves changing the basis as follows:
\begin{align}
    e^\mu_{(a)} = \tensor{\Lambda}{^{(b)}_{(a)}} \partial^\mu_{(b)}\,.
\end{align}
Here, \(\tensor{\Lambda}{^{(b)}_{(a)}}\) represents the coefficients of the fluid frame in the BL coordinate system.
After solving the equation, it is crucial to link the obtained result, \(\mathcal{S}^{\alpha\beta}\), with the Stokes parameters read by the observer. To achieve this, we must establish a local frame of reference at the observer's location. In line with our previous work \cite{Hu:2020usx}, we choose a zero angular momentum observer (ZAMO). However, our approach differs in that we adopt the `looking outward' gauge to facilitate the study of polarization as shown in Fig. \ref{fig:IAUconvent}. Consequently, the frame is established as follows:
\begin{align}
	e_{T}=\frac{g_{\phi\phi}\partial_t-g_{t\phi }\partial_{\phi}}{\sqrt{g_{\phi\phi}\left(g_{t\phi }^2-g_{\phi\phi}g_{tt}\right)}}\,,\quad\quad 
    e_{X}=\frac{\partial_\phi}{\sqrt{g_{\phi\phi}}}\,,\quad\quad 
    e_{Y}=-\frac{\partial_\theta}{\sqrt{g_{\theta\theta}}}\,,\quad \quad 
    e_{Z}=\frac{\partial_r}{\sqrt{g_{rr}}}
    \,.
\end{align}
In our camera model, for an observer situated far from the black hole, the field of view angle is typically quite small. Consequently, we can project the polarization tensor of each pixel onto a common set of axes to derive the corresponding Stokes parameters. By employing this projection coordinate system, we can map the values of the Stokes parameters onto the screen. These Stokes parameters are defined in accordance with IAU standards. At this stage, on the imaging plane \((e_{X}, e_{Y})\), the electric-vector position angle (EVPA) is given by
\begin{align}
    \chi=\dfrac{1}{2}\arctan\pa{\dfrac{U}{Q}}\,,
\end{align}
and the linear polarization vector can be expressed by the following equation:
\begin{align}
\vec{P} = \dfrac{\sqrt{Q^2 + U^2}}{I}\left(\cos\chi\,,\sin\chi\right)\,.
\end{align}

At this point, let us summarize the entire procedure for obtaining the polarized image and Stokes Parameters using \codename:

1. \textbf{Backward Ray-Tracing:} Initially, we employ the backward ray-tracing technique to trace null geodesics from the observer's standpoint backward in time. This allows us to determine the endpoints of the light rays and identify whether they are captured by the black hole or continue towards infinity. (In practice, an outer boundary, such as \( r = 500 \), is set far from the black hole.)

2. \textbf{Forward Progression and Radiation Transfer:} Starting from the endpoints identified in the previous step, we then trace the light rays forward in time while simultaneously solving the radiation transfer equations (\ref{eqn:fszyfcnew0}) and (\ref{eqn:fszyfcnew}).

3. \textbf{Projection and Stokes Parameters:} We continue this process until the light rays return to the observer. At this point, we project the polarization tensor onto the observer's screen, thereby derive the Stokes parameters as read by the observer.

\section{Numerical scheme and code verification}\label{sec3}

In this section, we aim to validate the accuracy of our publicly available covariant polarized radiative transfer code, \codename, by comparing its output with the results from various established schemes. Our validation process is divided into four segments: the validation against analytical solutions, the thin-disk validation, the analytical thick-disk validation, and  the validation of thick-disk models generated by GRMHD simulations. In our computations, we use a Kerr black hole with a spin parameter \( a = 0.94 \). It is important to note that our code is versatile and can be applied to any spacetime. In the thin and thick disk validations, we use radiative coefficients derived from the thermal distribution of electrons, as detailed in Appendix \ref{sec:FC}. For the observing frequency, we choose \(\nu_o = 230\, \text{GHz}\), which corresponds to the actual observation frequency of the EHT.

\subsection{Validation of analytical results}

Our first test problem involves validating the results of the constant-coefficient non-relativistic polarized transport equation, with its analytical solution outlined in Appendix C of \cite{Dexter:2016cdk}. This examination aims to assess the precision of our chosen numerical integrator. It is important to note that the two tests conducted in this paper have been independently verified in previous studies \cite{Dexter:2016cdk, Moscibrodzka:2017lcu, Bronzwaer:2020kle, EventHorizonTelescope:2023hqy}. Our settings are consistent with these references, where the initial conditions for both tests are set as \(I = Q = U = V = 0\) and the integration interval is selected as \(\lambda \in (0, 3)\). 

\begin{table}[ht]
    \centering
    \begin{tabular}{|c|cccc|cccc|ccc|}
    \hline
    Test & $j_I$ & $j_Q$ & $j_U$ & $j_V$
    & $a_I$ & $a_Q$ & $a_U$ & $a_V$
    & $r_Q$ & $r_U$ & $r_V$ 
      \\ \hline 
    Emission/Absorption &$2$& $1$ &$0$ &$0$ &$1$ &$1.2$ &$0$ &$0$ &$0$ &$0$ &$0$    \\
    \hline
    Rotation  &$0$& $0.1$ &$0.1$ &$0.1$ &$0$ &$0$ &$0$ &$0$ &$10$ &$0$ &$-4$    \\
    \hline
    \end{tabular}
    \caption{The selection of parameters in the validation of analytical results.}
    \label{table:CS}
\end{table}

\begin{figure}[ht]
    \centering
    \includegraphics[width=6in]{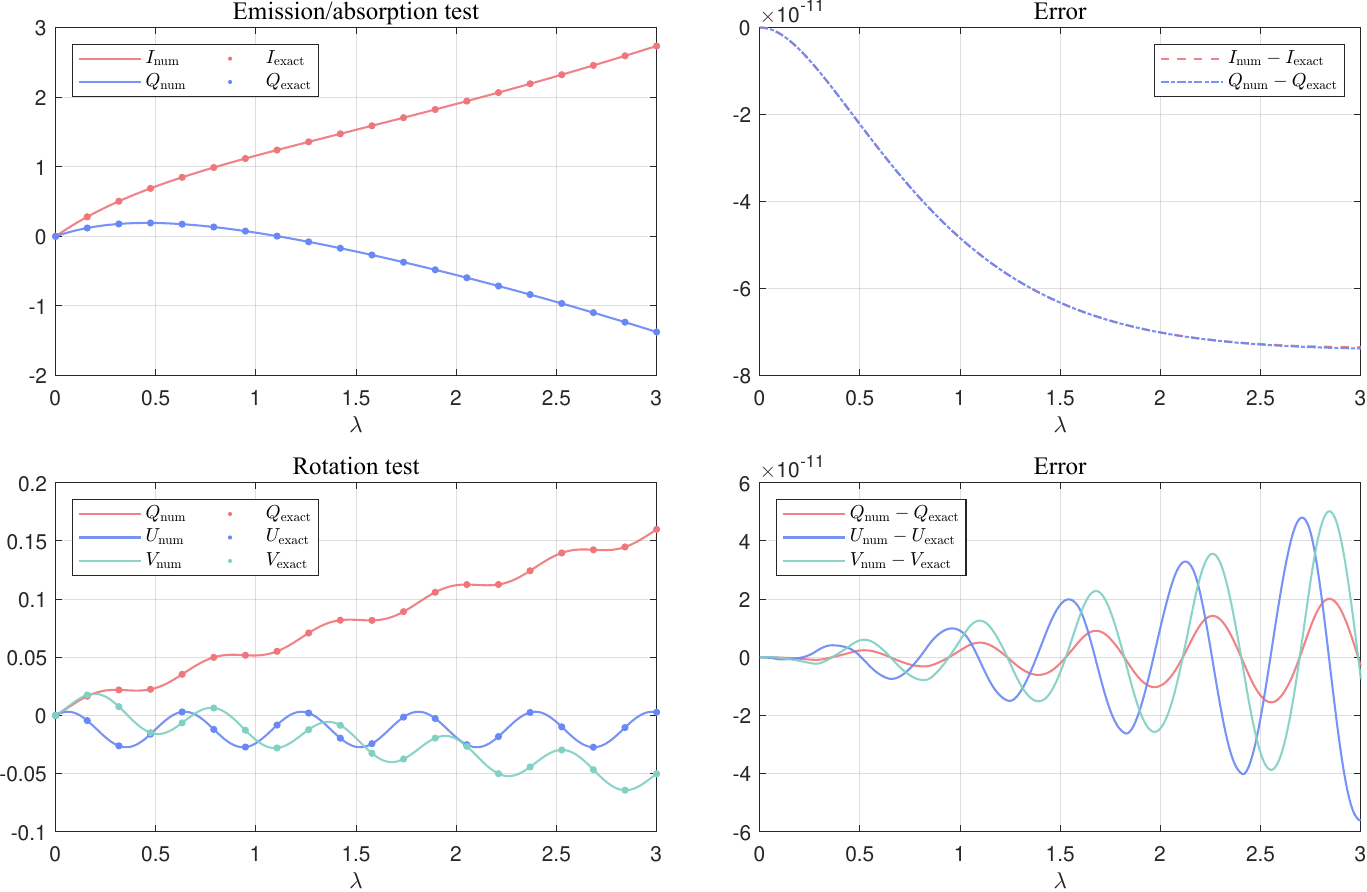}
    \caption{The first row corresponds to the outcomes derived from the emission/absorption examination, while the second row delineates the findings from the rotation test. The left column illustrates the evolution of Stokes parameters with respect to the affine parameter \(\lambda\), where the quantities not shown in the plot remain consistently at zero. The right column depicts the absolute discrepancy between the analytical solution and the numerical solution obtained using \codename.}
    \label{fig:anyresult}
\end{figure}

To simplify the representation of the analytical solution, we analyze the effects of  emission,  absorption, and  Faraday rotation separately. The coefficients used in the tests are presented in Table \ref{table:CS}. In the first test, we consider only the impacts of emission and absorption, setting all the Faraday rotation coefficients to zero. In the second test, we account solely for the effects of emission and Faraday rotation, with all the absorption coefficients set to zero.

In \codename, we employ the 4th-order Runge-Kutta (RK4) method to tackle this issue, setting the absolute error tolerance at \( \bar{\epsilon}=1 \times 10^{-12} \). Fig. \ref{fig:anyresult} illustrates the numerical outcomes alongside the analytical solutions for two distinct assessments, as well as the absolute discrepancies between the numerical and analytical results. From the comparison between the numerical results and the analytical outcomes, it is evident that the numerical results provided by \codename\, are reliable. The examination of the absolute error reveals that the solver's results align well with our designated error tolerance.

\subsection{Validation of thin disc}

In the second segment, we aim to test the accuracy of \codename\, in computing the parallel transport of polarization vectors. For this evaluation, we consider the imaging of a luminous, opaque thin disc positioned at the equatorial plane, with the rest of the space assumed to be a vacuum. Given that the disc is opaque, our focus is solely on its primary image. We will conduct calculations using two distinct methods: one employing our \codename\, program and the other utilizing the Penrose-Walker (PW) constant for computation\footnote{The computational details for imaging the thin disc using the PW constant are provided in App. \ref{BB}.}. For this assessment, we assume that the results obtained through the PW constant are entirely precise. By comparing the disparities between the outcomes calculated in these two methods, we can validate the precision of \codename\, in determining the parallel transport of polarization vectors.

\subsubsection{Parameter configuration}

In the thin-disc test model under consideration, the light rays propagate solely through a vacuum. Consequently, throughout the evolution equations, the coefficients related to emission, absorption, and Faraday rotation—denoted by \( j_S, a_S, r_S \) for \(\{I, Q, U, V\} \)—all vanish. In this scenario, the initial Stokes parameters are determined by the emission of photons at the first intersection point on the equatorial plane, traced back along the light ray from the camera. For simplicity, we set the initial Stokes parameters to be proportional to the instantaneous emission coefficient, i.e., \( \mathcal{S}_s \propto j_S/\nu^2 \).

To compute the emission coefficient on the equatorial plane, we need to specify the physical parameters of the accretion disk at this plane: the electron number density \( n_e \), the electron temperature \( T_e \), and the magnetic field \( B^\mu \). Based on the power-law relationships derived from GRMHD simulations and fitted in \cite{Yuan:2003dc, Broderick:2005at}, we adopt the following distributions for the physical parameters of the thin disk:
\begin{align}
n_e &= n_0 \left(\dfrac{r_+}{r}\right)^{1.1}\,,\nn \\
T_e &= T_0 \left(\dfrac{r_+}{r}\right)^{0.84}\,, \nn\\
B^2 &= B^2_0 \left(\dfrac{r_+}{r}\right)\,,
\end{align}
where \( n_0 \), \( T_0 \), and \( B_0 \) represent the values of the accretion disk at the event horizon. Drawing insights from the observations of M87* and the results from GRMHD simulations, we choose \( n_0 = 10^6 \, \text{cm}^{-3} \), \( T_0 = 10^{12} \, \text{K} \), and \( B_0 = 10 \, \text{Gauss} \) as the reference values.

\begin{figure}[ht]
    \begin{center}
        \includegraphics[width=6in]{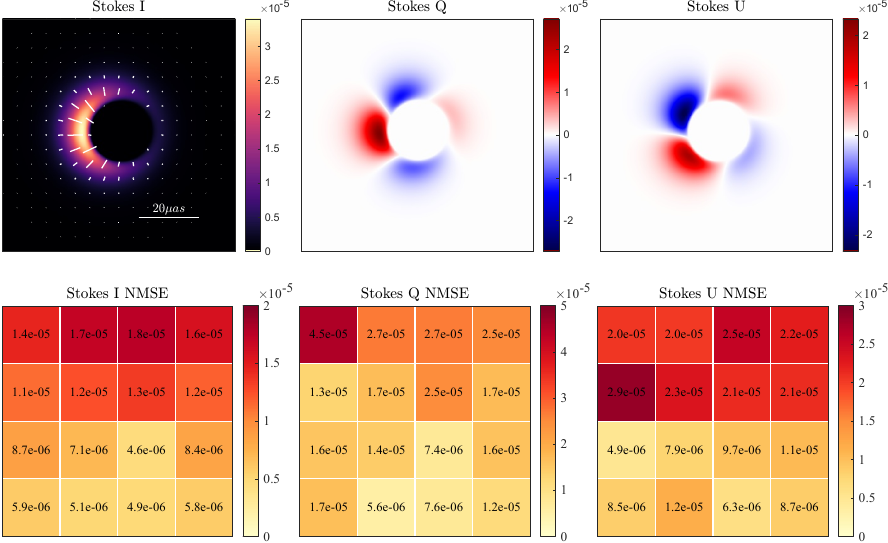}
    \end{center}
    \caption{The results of the thin disk test are presented in two rows. The first row shows images of Stokes $I$, $Q$, and $U$ generated using \codename. The second row features a plot of the NMSE, calculated by comparing the results from \codename\, with PW-constant computations across different regions.}
    \label{fig:thindisk}
\end{figure}

In modeling the motion of the accretion disk fluid, we consider a simplified flow pattern. We assume that outside the innermost stable circular orbit (ISCO), the fluid follows Keplerian motion. The angular velocity of the fluid at radius \( r \) is given by
\begin{align}
\frac{u^\phi}{u^t} = \frac{1}{r^{3/2} + a}\,,
\end{align}
where \( a \) is the spin parameter. Within the ISCO, the fluid transitions to a plunging motion. In this region, the fluid's energy \( E \) and angular momentum \( L \) are equal to those of the fluid in Keplerian motion at the ISCO. The choice of magnetic field configuration is somewhat arbitrary. In this study, we adopt a straightforward approach by selecting a purely azimuthal magnetic field distribution, that is, 
\begin{align}
    B^\mu\pa{r}\propto \sqrt{\dfrac{1}{r g_{\phi\phi}}}\delta_\phi^\mu \,.
\end{align}

\subsubsection{Results}

We adopt the evaluation metric utilized in the article \cite{EventHorizonTelescope:2023hqy}, specifically employing the normalized mean squared error (NMSE) to assess outcomes. The NMSE allows us to gauge the similarity between two images:
\begin{align}
\text{NMSE}(A,B) = \frac{\sum_{i,j} \left|A(i,j) - B(i,j)\right|^2}{\sum_{i,j} \left|A(i,j)\right|^2}\,.
\end{align}
In this equation, \(A(i,j)\) and \(B(i,j)\) represent the values of a particular Stokes parameter at pixel \((i,j)\) in the two images, respectively.

Here, we select an observer positioned at \( r_o = 600 \), \( \theta_o = 17^\circ \), with a field of view of \( \pi/96 \). The imaging pixel density is \( 2048 \times 2048 \), and the generated results for Stokes $I$, $Q$, and $U$ can be observed in the first row of Figure \ref{fig:thindisk}. Since the circular polarization emission is not considered in the thin-disk test, only the results for $I$, $Q$, and $U$ are provided here.

We divide the images generated by both methods into 16 subregions, each consisting of \( 512 \times 512 \) pixels. Subsequently, we calculate the NMSE for each subregion individually. The results are depicted in the second row of Figure \ref{fig:thindisk}. It is evident that for Stokes $I$, $Q$, and $U$, \codename\, demonstrates low NMSE values compared to the exact results, with maximum values on the order of \( 10^{-5} \). This suggests a high level of consistency between the results obtained using \codename\, and the accurate results. This sufficiently indicates that our program can maintain a high level of accuracy in terms of parallel transport.

\subsection{Validation of the Thick Disk Model}\label{sec:VThickDM}

Next, we focus on thoroughly validating the thick disk by considering both the parallel transport of polarization and the interaction of light with the plasma. For the validation of the thick disk, we require a plasma model, and here we employ the analytical magnetofluid model proposed in \cite{Hou:2023bep, Zhang:2024lsf}.

\subsubsection{Parameter Configuration}

\begin{figure}[h]
    \centering 
    \includegraphics[width=5in]{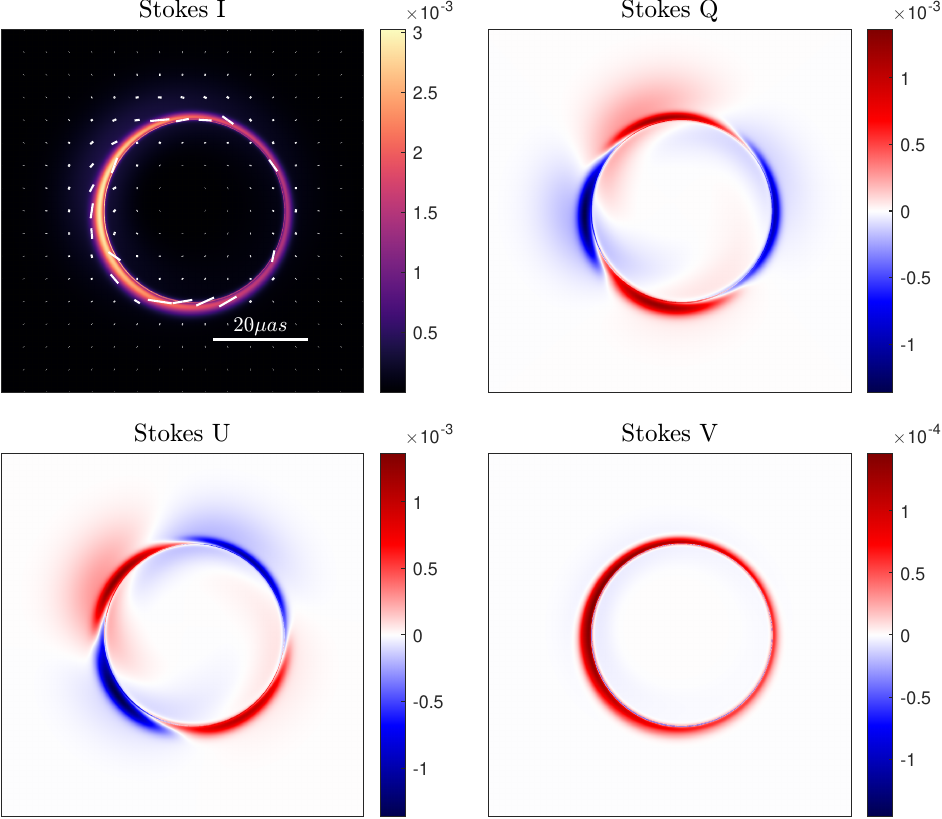}
    \caption{The resulting Stokes parameters \( I, Q, U, V \) are generated by the \codename\,. The pixel density is 1024 \(\times\) 1024.}
    \label{fig:hou1}
\end{figure}

\begin{figure}[h]
    \centering 
    \includegraphics[width=6in]{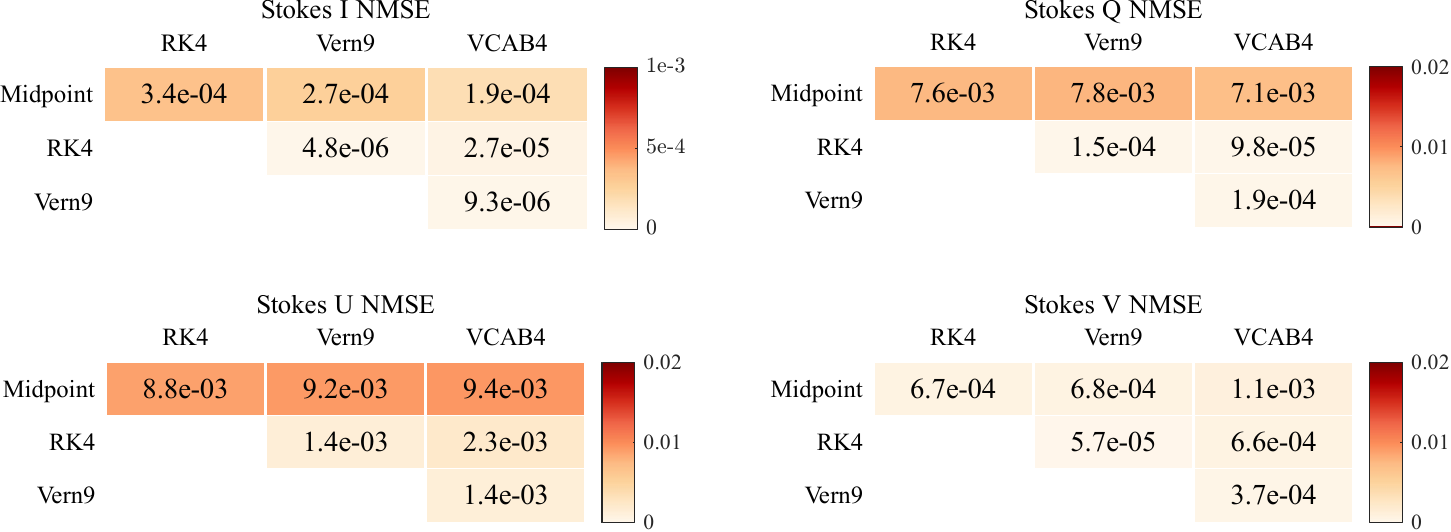}
    \caption{Comparison of the normalized mean squared error (NMSE) of the results generated using different numerical methods in \codename.To facilitate a more intuitive understanding, we will apply shading to the cells in the table.}
    \label{fig:hou2}
\end{figure}

In this model, the accretion flow is considered as a ballistic approximation of axisymmetric spherically symmetric accretion, with the fluid moving along geodesics and satisfying constancy of $\theta$. The four-velocity of this fluid can then be represented as
\begin{align}
    \begin{array}{ll}
         u^r=\sigma_r \dfrac{\sqrt{R}}{\Sigma}\,,
         \qquad
         \qquad
      &  u^t=E\br{1+\dfrac{2r\pa{r^2+a^2}}{\Delta\Sigma}}-
        \dfrac{2arL}{\Delta\Sigma}\,,
       \\
        u^\theta=0\,,
        &u^\phi=E\dfrac{2ar}{\Delta\Sigma}+
        L\dfrac{\csc^2\theta\pa{\Delta-a^2\sin^2\theta}}{\Delta\Sigma}\,.
    \end{array}
\end{align}
The angular momentum \( L =\sigma_L a \sqrt{E^2-1} \sin^2\theta \) and the Carter constant \( Q = -a^2 (E^2-1) \cos^4\theta \) appearing in the above equation are functions of the energy \( E \) and the polar angle \( \theta \). The radial potential can be expressed as:
\begin{align}
R(r,\theta) = (E^2-1)r^4 + 2r^3 + 2a^2(E^2-1)\cos^2\theta \, r^2 + 2((aE-L)^2+Q)r - aQ^2,
\end{align}
where we use \( \sigma_r = \pm 1 \) to indicate the direction of fluid motion along the radial axis and \( \sigma_L = \pm 1 \) to represent the direction along the azimuthal axis.In this study, we define \(\sigma_r = -1\) to represent an ingoing accretion flow and \(\sigma_L = 1\) to indicate a retrograde accretion flow. For simplicity in subsequent program validation, we will set \( E = 1 \).

The electron number-density distribution and electron temperature distribution of the accretion flow can be uniquely determined by the distribution of the accretion flow at the black hole event horizon, given by:
\begin{align}
n_e(r,\theta) = n_e(r_+,\theta) \sqrt{\dfrac{R(r_+,\theta)}{R(r,\theta)}}\,,\quad
T_e(r,\theta) = T_e(r_+,\theta) \left(\dfrac{R(r_+,\theta)}{R(r,\theta)}\right)^{\frac{1+z}{3(2+z)}}\,.
\end{align}
Here, we set the density distribution \( n_e(r_+,\theta) \) at the horizon as a Gaussian distribution and the electron temperature distribution \( T_e(r_+,\theta) \) as a constant:
\begin{align}
n_e(r_+,\theta) = n_0 \exp\left[-\left(\dfrac{\sin\theta - \sin\theta_J}{\sigma}\right)^2\right]\,,\quad T_e(r_+,\theta) = T_0\,,
\end{align}
where the distribution parameters are chosen as \( \theta_J = \pi/2 - 10^{-3} \), \( \sigma = 0.2 \), and \( z = 20 \). Similar to the thin disk test, we refer to the observational values of M87* and the simulation results of GRMHD to set \( n_0 = 10^6 \text{ cm}^{-3} \) and \( T_0 = 10^{12} \text{ K} \).
Regarding the magnetic field configuration, we opt for the simplest split monopole configuration, given by
\[
B^\mu(r,\theta) = -\text{sign}(\cos\theta) \dfrac{\Psi_0}{\Sigma u^r} (u_t u^\mu + \delta^\mu_t),
\]
where \(\Psi_0 > 0\) is a constant, uniquely determined by the magnetic field strength at the event horizon, \(\left|B(r_+,\theta_J)\right| = B_0 = 10 \text{ Gauss}\).

\subsubsection{Results}

In the thick disk tests, we employ two distinct validation approaches. The first approach involves cross-validating results by using different numerical methods with varying convergence accuracies in \codename. The second approach, referred to as the RAPTOR-like code\footnote{Here, we are not directly using the RAPTOR code. Instead, we have independently developed a program for validation based on the theoretical framework of the RAPTOR code as outlined in \cite{Bronzwaer:2020kle}. This custom program, referred to as the RAPTOR-like code, serves as our means of verification.}, involves the independent calculations of parallel transport of polarized vectors and the interaction with the plasma, similar to the RAPTOR code \cite{Bronzwaer:2020kle}. In contrast, \codename\, integrates both effects simultaneously in its computations. The results are then compared against those generated by \codename. In both validation sets, we maintain consistency with the thin-disk tests by selecting an observer positioned at \( r_o = 600 \), \( \theta_o = 17^\circ \), and a field of view angle of \( \text{fov} = \pi/96 \). Due to computational time constraints, we opt for a pixel density of \( 128 \times 128 \) in subsequent tests.

\begin{table}[h]
\centering
\begin{tabular}{|c|c|c|c|}
\hline
NMSE I & NMSE Q & NMSE U & NMSE V \\
\hline
$3.4\times 10^{-4}$ & $1.4\times 10^{-2}$ & $5.7\times 10^{-3}$ & $1.2\times 10^{-2}$ \\
\hline
\end{tabular}
\caption{\codename\, vs. RAPTOR-like.}
\label{tab:Coport vs. RAPTOR-like}
\end{table}

The initial segment involves the validation within the \codename\, program, where two distinct numerical solution approaches are employed: explicit Runge-Kutta methods and Adams-Bashforth explicit methods. For the Runge-Kutta (RK) methods, we utilize three tiers of algorithms: the second-order midpoint method (Midpoint), the canonical fourth-order Runge-Kutta method (RK4), and Verner's high-order ``Most Efficient" 9/8 Runge-Kutta method (Vern9). For the Adams-Bashforth (AB) methods, we implement the fourth-order Adams method (VCAB4). To ensure minimal disparities in computational time, we set the absolute tolerance values as follows: $10^{-5}$ for Midpoint,$10^{-7}$ for RK4, $10^{-9}$ for Vern9, and $10^{-8}$ for VCAB4.

As the differences between the results obtained using various numerical methods in \codename\, are imperceptible to the naked eye, we exclusively present the high-definition results generated by the RK4 method. These visual representations can be found in Fig. \ref{fig:hou1}. Additionally, Fig. \ref{fig:hou2} highlights the disparities between the results produced by different numerical methods within \codename. It is evident that these methods yield the results with minimal deviations across different numerical precisions, indicating both the convergence of results in \codename\, and the limited impact of numerical-method selection on the outcomes.

\begin{figure}[h]
    \centering 
    \includegraphics[width=5in]{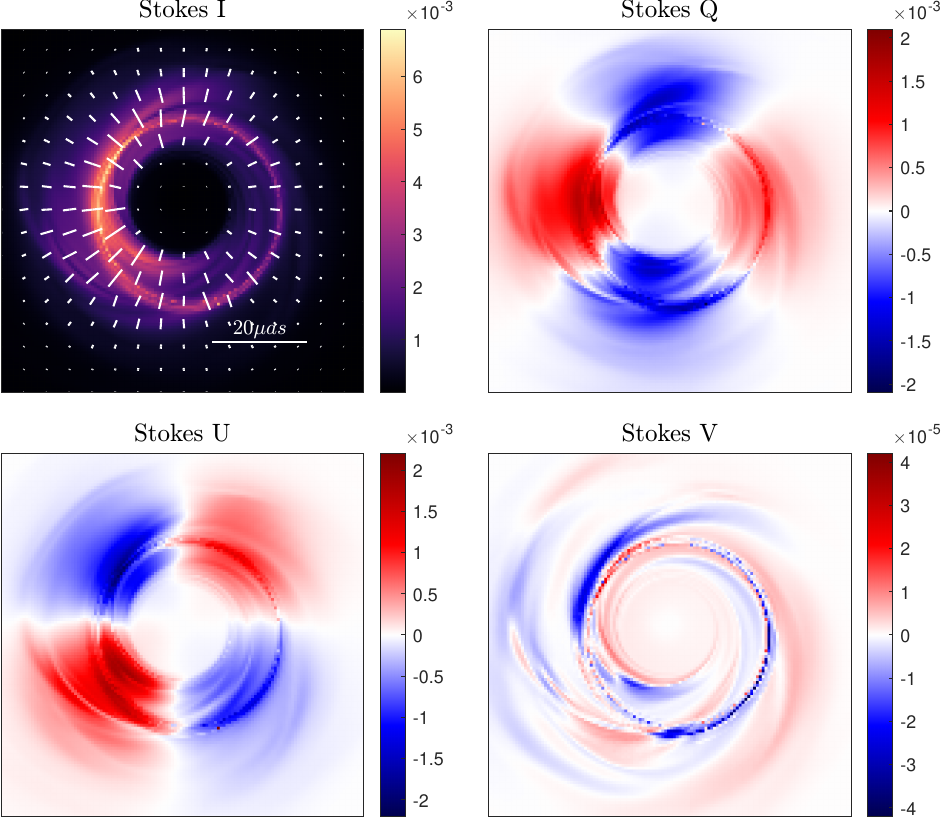}
    \caption{The resulting Stokes parameters \( I, Q, U, V \) of the GRMHD snapshot are generated by the \codename\,. The pixel density is 128 \(\times\) 128.}
    \label{fig:GRMHD}
\end{figure}

\begin{table}[h]
    \centering
    \begin{tabular}{|c|c|c|c|}
    \hline
    NMSE I & NMSE Q & NMSE U & NMSE V \\
    \hline
    $3.8\times 10^{-3}$ & $2.4\times 10^{-2}$ & $1.8\times 10^{-2}$ & $4.9\times 10^{-2}$ \\
    \hline
    \end{tabular}
    \caption{\codename\, vs. RAPTOR-like.}
    \label{tab:GRMHD Coport vs. RAPTOR-like}
\end{table}

The second segment involves a comparative validation between \codename\ and the RAPTOR-like method, which utilize distinct computational principles. In this analysis, we maintain a fixed step size of \(\Delta\lambda=0.02\) for the RAPTOR-like code. We also compute the NMSE results between \codename\ and RAPTOR-like, as presented in Table \ref{tab:Coport vs. RAPTOR-like}. From the table, it can be observed that the highest order of NMSE is on the scale of only \(10^{-2}\), indicating consistency with the NMSE magnitudes reported in the GRMHD snapshot test outputs as described in \cite{EventHorizonTelescope:2023hqy}. The alignment of NMSE magnitudes across different code outputs reaffirms the accuracy of the results obtained by \codename, despite differing in computational principles, converges in results. This convergence thereby validates the integrity of \codename's outcomes.

\subsection{GRMHD Snapshot}

Furthermore, \codename\ has developed an interface to visualize the numerical results obtained from GRMHD simulations. We utilize HARM \cite{Gammie:2003rj}, an open-source GRMHD tool, to simulate the accretion process near a supermassive black hole. Subsequently, \codename\ generates the corresponding polarization images. Since the data obtained through HARM is used solely for validating our code, we will refrain from discussing the physical interpretation of the HARM results in this context\footnote{The HARM data is publicly accessible at the following URL: https://github.com/JieweiHuang/Coport-Data.}. Additionally, we use our self-developed RAPTOR-like code to image the data obtained from the same HARM simulations, thereby cross-validating the accuracy of our approach. 

The GRMHD test configuration mirrors that of Sec. \ref{sec:VThickDM}, where we maintain a fixed step size of \(\lambda = 0.02\) for the RAPTOR-like program. For the \codename\ algorithm, we employ the RK4 method with an absolute tolerance of \(10^{-7}\). Concerning the camera settings, we opt for an observer situated at \(r_o = 600\), \(\theta_o = 17^\circ\), and a field of view angle of \(\pi/96\).

The outcomes generated by \codename\ are presented in Fig. \ref{fig:GRMHD}. Additionally, the disparities between the results produced by the \codename\ algorithm and the RAPTOR-like code are outlined in Table \ref{tab:GRMHD Coport vs. RAPTOR-like}. Table \ref{tab:GRMHD Coport vs. RAPTOR-like} shows that the NMSE values are on the order of \(10^{-2}\). This again indicates consistency with the NMSE magnitudes reported in the GRMHD snapshot test outputs, as described in \cite{EventHorizonTelescope:2023hqy}. Particularly noteworthy is the significant variance in Stokes $V$, which we attribute to the RAPTOR-like code's inadequately short step length, resulting in pronounced deviations\footnote{We developed the RAPTOR-like code specifically to validate Coport, and therefore, we did not optimize it for better performance. As a result, its computational speed is relatively slower. Consequently, we chose not to further reduce the step size for a stronger validation.}.

\section{Summary}\label{sec4}

To improve our understanding of polarized observations in the environments with strong gravitational fields, we have developed a highly efficient algorithm named \codename\ for polarized radiative transfer across various spacetime frameworks. In contrast to the existing schemes such as ipole \cite{Moscibrodzka:2017lcu} and RAPTOR \cite{Bronzwaer:2020kle}, our implementation streamlines the calculation process by removing the need to separate the contributions from gravitational Faraday effect and plasma interaction. Instead, we address both gravitational effects and plasma interactions concurrently by solving the covariant polarized radiative transfer equation directly. Our formalism not only adeptly encapsulates the covariance inherent in the gravitational theory but also ensures high efficiency in numerical computations.

We have validated the accuracy and the precision of our new algorithms through multiple methods: comparing them with analytical solutions and validating them using both thin-disk and thick-disk models. For the most intricate thick-disk models, our validation process involves two steps. First, we cross-validated outcomes using diverse numerical methodologies with varying levels of convergence accuracy within our code. Additionally, we compared these results with those obtained using the RAPTOR-like algorithms. This approach not only confirms the convergence of outcomes within our code but also demonstrates that the choice of numerical methods has a negligible impact on the final results. Consequently, this affirms the robustness of our algorithm's outputs.

The polarization outcomes of the plasma surrounding black holes are highly significant, not only for astronomical observations but also for theoretical studies of the plasma and the validation of various gravitational theories. \codename\ is a novel instrument designed to analyze the radiative properties of diverse plasma models in arbitrary spacetimes. These simulations aim to align with current and future observations of polarized radiation from relativistic plasma around black holes, neutron stars, and potentially other intriguing celestial objects.

\section*{Acknowledgments}
We are deeply grateful to Zhen Zhong for the invaluable suggestions regarding Julia, as well as for the assistance with compiling and executing the code. The work is partly supported by NSFC Grant Nos. 12205013 and 12275004. MG is also endorsed by "the Fundamental Research Funds for the Central Universities" with Grant No. 2021NTST13. 

\appendix
\label{app}
\section{Covariant polarized radiative transfer equation}\label{sec:A}

In the section, our objective is to reassess the formalism for covariant polarized radiative transport as delineated in  \cite{2012ApJ...752..123G}. This overview closely aligns with the content of \cite{2012ApJ...752..123G}, with the primary goal being the rectification of notation. Additionally, we aim to include nontrivial calculation steps that were overlooked in the original work \cite{2012ApJ...752..123G}.

\subsection{A single optical ray with a specific polarization state}

Within the framework of general relativity, the Maxwell equations are formulated as:
\begin{align}
    \nabla^\nu \nabla_\nu A^\mu+R^\mu_{\ \  \nu}A^\nu=4\pi J^\mu\,.
    \label{eqn:Maxwelleqn}
\end{align}
Here, we have utilized the Lorentz gauge $\nabla_\mu A^\mu=0$, $A^\mu$ represents the potential, $R_{\mu\nu}$ denotes the Ricci tensor, and $J^\mu$ signifies the current generated by charged particles. For a light ray with a specific polarization state, the WKB approximation can be employed by expanding the potential $A^\mu$ in terms of a small parameter $\epsilon$,
\begin{align}
    A^\mu=
\left(a^\mu+\epsilon b^\mu+\cdots \right)\exp
\left({i\dfrac{\theta}{\epsilon}}\right)\,,
\label{eqn:WKBform}
\end{align}
We propose that the current \( J^\mu \) follows a linear response law, expressed as \( J^\mu = \dfrac{1}{\epsilon} \Pi^{\mu\nu} A_\nu \), where \( \Pi^{\mu\nu} \) represents the linear response tensor. By substituting the expression for $J^\mu$ along with Equation (\ref{eqn:WKBform}) into Equation (\ref{eqn:Maxwelleqn}) and considering the first two orders of the equation, we obtain 
\begin{align}
&\mathcal{O}\left(\dfrac{1}{\epsilon^2}\right):\qquad
k_\mu k^\mu=0\,,\label{geoe}\\
&\mathcal{O}\left(\dfrac{1}{\epsilon}\right):\qquad
k^\rho\nabla_\rho a^\mu+\dfrac{1}{2}a^\mu \nabla_\rho k^\rho
=2\pi i \Pi^{\mu\rho}a_\rho\,,\label{sece}
\end{align}
where $k_\mu=\partial_\mu\theta$. In reality, Eq. (\ref{geoe}) can be expressed as $k^\nu\nabla_\nu k^\mu=0$, representing the geodesic equation governing the paths of light rays within the framework of geometric optics approximation.  By multiplying both sides of Eq. (\ref{sece}) by $\bar{a}^\nu$ and subsequently summing after conjugating the entire equation, we derive
\begin{align}
\nabla_\rho\left(
k^\rho a^\mu \bar{a}^{\nu}
\right)=H^{\mu\nu\rho\sigma}a_\rho \bar{a}_\sigma\,,
\label{eqn:gxejgx}
\end{align}
where, we have introduced a new tensor following \cite{2012ApJ...752..123G},
\begin{align}
    \displaystyle{H^{\mu\nu\rho\sigma}=2\pi i \left(g^{\nu\sigma}\Pi^{\mu\rho}-g^{\mu\rho}\bar{\Pi}^{\nu\sigma}\right)}\,.
\end{align}
\par 
Based on \(\Pi^{\mu\rho}\),we can define the Hermitian and anti-Hermitian tensors as follows:
\bea
\mR^{\alpha\beta}/i = 2\pi  \left(\Pi^{\alpha\beta} + \bar{\Pi}^{\beta\alpha}\right)\,, \quad
\mA^{\alpha\beta}/i = 2\pi  \left(\Pi^{\alpha\beta} - \bar{\Pi}^{\beta\alpha}\right)\,,
\eea
where \(\mR^{\alpha\beta}/i\) is the Hermitian tensor that conserves total energy, and \(\mA^{\alpha\beta}/i\) is the anti-Hermitian tensor that induces dissipation. One can then define the following tensors
\bea
\mR^{\mu\nu\rho\sigma} = \dfrac{1}{2}\left(g^{\nu\sigma}\mR^{\mu\rho} - g^{\mu\rho}\mR^{\sigma\nu}\right)\,, \quad \mA^{\mu\nu\rho\sigma} = \dfrac{1}{2}\left(g^{\nu\sigma}\mA^{\mu\rho} + g^{\mu\rho}\mA^{\sigma\nu}\right)\,,
\eea
and thus we have
\bea
H^{\mu\nu\rho\sigma} = \mR^{\mu\nu\rho\sigma} + \mA^{\mu\nu\rho\sigma}\,,\label{hdec}
\eea
where \(\mR^{\mu\nu\rho\sigma}\) contains the non-dissipative terms describing the generalized Faraday rotation part, and \(\mA^{\mu\nu\rho\sigma}\) contains the dissipative terms, representing the absorption part. Clearly, in vacuum, we have $H^{\mu\nu\rho\sigma}=0$, hence the equation in vacuum reduces to $\nabla_\rho\left(k^\rho a^\mu \bar{a}^{\nu}\right)=0$.

At this point, we consider the average energy-momentum tensor of a monochromatic electromagnetic wave, given by
\begin{align}
T^{\mu\nu}=\dfrac{1}{4\pi}
\langle  \Re\{ F^\mu_{\ \ \rho}\}
\Re\{ F^{\nu\rho}\}-\dfrac{1}{4}g^{\mu\nu}\Re\{ F_{\rho\sigma}\}
\Re\{ F^{\rho\sigma}\}\rangle=\dfrac{1}{8\pi}\lrg{a_\rho \bar{a}^{\rho}}k^\mu k^\nu\,.
\end{align}
Herein, we use `$\langle\cdot\rangle$' to represent the time-average over one period for `$\cdot$'. Monochromatic electromagnetic waves can exist in various polarization states. In the case of a photon with a definite polarization vector, we can choose an arbitrary direction's polarization basis vector, denoted as $\epsilon_\alpha$, to characterize its polarization properties. We then extend the average energy-momentum tensor of monochromatic electromagnetic waves to the average energy-momentum tensor of polarized monochromatic electromagnetic waves, denoted as:
\begin{align}
    \pa{T^{\mu\nu}}^{\alpha\beta}=\dfrac{1}{8\pi}\lrg{a^\alpha \bar{a}^{\beta}}k^\mu k^\nu\,.
    \label{eqn:ndzljhbs}
\end{align}
Therefore, $\pa{T^{\mu\nu}}^{\alpha\beta}\epsilon_\alpha\bar{\epsilon}_\beta$ yields the average energy-momentum tensor $T^{\mu\nu}$ of monochromatic electromagnetic waves in the specified polarization state.

\subsection{Photon tube and polarization tensor}

Next, we extend the trajectory of an individual photon to that of a group of photons, at which point rays are generalized to a `photon tube'. For a group of photons, we can statistically analyze the energy-momentum tensor of the electromagnetic waves. By choosing a local rest observer $u^\mu$ at the spacetime point $x^\mu$, we can establish three spatial frames. subsequently, for the average energy-momentum tensor of all electromagnetic waves with a particular polarization state, we can represent it using a distribution function as follows:
\begin{align}
    T^{\mu\nu}(x)=
\int f(x,p)p^\mu p^\nu \dif V_p\,.
\label{eqn:ndzlfbhsbs}
\end{align}  
Here, $f(x,p)$ is the photon distribution function, which specifies the total count of photons in the quantum state corresponding to the phase space element $\dif V_x \dif V_p=\dif^3x\dif^3p$, as follows 
\bea
\dif N(x,p)=f(x,p)\dif^3x\dif^3p\,.
\eea  
It is important to note that since the spatial volume element $\dif V_x=\sqrt{-g}p^0\dif^3x=\sqrt{-g}p^0\dif x^1\dif x^2\dif x^3$ and 
 momentum space volume element $\dif V_p=\dfrac{\dif^3p}{\sqrt{-g}p^0}=\dfrac{\dif p_1\dif p_2\dif p_3}{\sqrt{-g}p^0}$, as well as the particle number $N$ are invariant, the scalar distribution function $f(x,p)$ is also invariant.

For a monochromatic wave with momentum $k^{\mu}=(k^0,\vec{k})$, its distribution function should only contribute at the point $\vec{p}=\vec{k}$ in momentum space. Therefore, the corresponding distribution function should be expressed as:
\begin{align}
 f(x,p)=\dfrac{\dif N(x)}{\dif^3x} \delta^{(3)}\left(\vec{p}-\vec{k}\right)\,.
\end{align}
Substituting this into Eq. (\ref{eqn:ndzlfbhsbs}), we have
\begin{align}
T^{\mu\nu}=\dfrac{\dif N(x)}{\sqrt{-g}{k^0} \dif^3x}{k^\mu k^\nu}\,.
\end{align} 
If we extend the average energy-momentum tensor of monochromatic electromagnetic waves to the average energy-momentum tensor of polarized monochromatic electromagnetic waves, we can similarly define
\begin{align}
\pa{T^{\mu\nu}}^{\alpha\beta}=
\int\dfrac{\dif N^{\alpha\beta}(x,p)}{\dif^3x\dif^3p}p^\mu p^\nu \dfrac{\dif^3p}{\sqrt{-g} p^0}\,,
\end{align}
and the distribution function polarization tensor \cite{Portsmouth:2004ee}
\begin{align}
    \displaystyle{f^{\alpha\beta}(x,p)=\dfrac{\dif N^{\alpha\beta}(x,p)}{\dif^3x\dif^3p}}\,. 
\end{align}
Clearly, \( f^{\alpha\beta} \) satisfies the condition that \( f^{\alpha\beta} \epsilon_\alpha \bar{\epsilon}_\beta d^3p \, d^3x \) is proportional to the number of photons in the phase space element passing through a polarizer, which is oriented to select polarization \( \epsilon_\alpha \), per unit time. In a similar manner, for polarized monochromatic waves, the average energy-momentum tensor can be expressed as
\begin{align}
    \pa{T^{\mu\nu}}^{\alpha\beta}=
    \dfrac{\dif N^{\alpha\beta}(x)}{\sqrt{-g}k^0\dif^3x}k^\mu k^\nu \,.
    \label{eqn:ndzltjbs}
\end{align}

\par
By comparing Eqs. (\ref{eqn:ndzljhbs}) and (\ref{eqn:ndzltjbs}), we can derive
\begin{align}
    \dfrac{\dif N^{\alpha\beta}(x)}{\sqrt{-g}k^0\dif^3x}=
    \dfrac{1}{8\pi }\lrg{a^\alpha_i \bar{a}^{\beta}_i}\,.
    \label{eqn:ddcbzyfc}
\end{align} 
Now, let's consider how the distribution function polarization tensor $f^{\alpha\beta}$ evolves with the photons. Consider the distribution pattern of \( f^{\alpha\beta} \) in the invariant phase volume \( \Delta V_p \) at the point \((x, p)\). We have
\begin{align}
    f^{\alpha\beta}=\dfrac{\dif N^{\alpha\beta}(x,p)}{\dif^3x\dif^3p}\approx 
    \dfrac{1}{\Delta V_p}\sum_i \dfrac{1}{\sqrt{-g}k^0}
    \dfrac{\dif N^{\alpha\beta}_i}{\dif^3x}\,,
\end{align}
where, $\displaystyle{\sum_i}$ represents the sum over all possible modes of vibration, which are strictly equal in the case of $\Delta V_p \to 0$. By substituting Eq. (\ref{eqn:ddcbzyfc}) into the above formula, the distribution function becomes equivalent to
\begin{align}
    f^{\alpha\beta}=\dfrac{1}{8\pi \Delta V_p}\sum_i \lrg{a^\alpha_i \bar{a}^{\beta}_i}\,.
\end{align}
Considering the motion of the distribution function polarization tensor $f^{\alpha\beta}$ for a group of photons, we use the affine parameter $\lambda$ for the photons and obtain
\begin{align}
    k^\mu\nabla_\mu f^{\alpha\beta}= \dfrac{1}{8\pi \Delta V_p}\sum_i\left[   -\lrg{a^\alpha_i \bar{a}^{\beta}_i}\dfrac{\dif}{\dif \lambda }
    \ln\pa{ \Delta V_p}
    +
    k^\mu\nabla_\mu\pa{\lrg{a^\alpha_i \bar{a}^{\beta}_i}}
    \right] \,.
\end{align} 
By applying the Eq. (\ref{eqn:gxejgx}) derived under the WKB approximation and substituting the second term on the right side of the above formula, we obtain
\begin{align}
    k^\mu\nabla_\mu f^{\alpha\beta}&=\dfrac{1}{8\pi \Delta V_p}\sum_i\left[
        \tensor{H}{^\alpha^\beta_\mu_\nu}
   \lrg{a_{i}^\mu
    \bar{a}_{i}^\nu}-\lrg{a_i^\alpha \bar{a}_i^{\beta}}\pa{ \nabla_\mu k^\mu+\dfrac{\dif}{\dif \lambda }\ln\pa{ \Delta V_p}
   }
    \right]\,.
\end{align}
Furthermore, using the continuity condition of the photon tube, $\nabla_\mu k^\mu =\dfrac{1}{\Delta V} \dfrac{\dif \Delta V}{\dif \lambda}$, we obtain
\begin{align}
    k^\mu\nabla_\mu f^{\alpha\beta}=H^{\alpha\beta\mu\nu}
    f_{\mu\nu}
    -f^{\alpha\beta}
    \dfrac{\dif}{\dif \lambda}\left[
        \ln\pa{\Delta V\Delta V_p}
    \right]\,.
\end{align}
Finally, we use Liouville's theorem, which states that the phase space volume element $\Delta V_x \Delta V_p$ remains unchanged during evolution, and ultimately obtain
\begin{align}
    k^\mu\nabla_\mu f^{\alpha\beta}=H^{\alpha\beta\mu\nu}f_{\mu\nu}\,.\label{eqn:fbhsydfc0}
\end{align}
Note that when deriving the above equation, we did not consider the light source. We now add a source term to the right side of Eq. \ref{eqn:fbhsydfc0} to obtain the complete equation
\begin{align}
    k^\mu\nabla_\mu f^{\alpha\beta}=J^{\alpha\beta}+H^{\alpha\beta\mu\nu}f_{\mu\nu}\,.\label{eqn:fbhsydfc}
\end{align} 

Similarly, we can generalize the preceding equation by introducing the concept of the polarization tensor as
\begin{align}
    \Mc S^{\alpha\beta}=\dfrac{2h^4}{c^2}{f^{\alpha\beta }}\,.
\end{align}
It is noteworthy that, compared to the polarization tensor defined in \cite{2012ApJ...752..123G}, an additional factor of two is present here. This adjustment will facilitate subsequent calculations. Similarly, when selecting any polarization state basis vector \(\epsilon_\alpha\), the expression \(\mathcal{S}^{\alpha\beta} \epsilon_\alpha \bar{\epsilon}_\beta\) yields twice the invariant specific intensity of polarized light $I_\nu/\nu^3$.

Given that the polarization tensor and the distribution function polarization tensor differ only by a coefficient, it follows from Eq. (\ref{eqn:fbhsydfc}) that the equation of motion for the polarization tensor should be as follows
\begin{align}
    k^\mu\nabla_\mu \Mc S^{\alpha\beta}=\Mc J^{\alpha\beta}+H^{\alpha\beta\mu\nu}\Mc S_{\mu\nu}\,.\label{eqn:stkszlydfc}
\end{align}
In practical calculations, it is often necessary to consider the projection of the polarization tensor onto the orthogonal subspaces defined by the observer \( u^\mu \) and the light ray \( k^\mu \). To facilitate this, we can define a projection operator
\begin{align}
    P^{\mu\nu}=g^{\mu\nu}+u^\mu u^\nu-e^\mu_{(k)}e^\nu_{(k)}=
    g^{\mu\nu}-\dfrac{k^\mu k^\nu}{\omega^2}+\dfrac{u^\mu k^\nu}{\omega}+\dfrac{k^\mu u^\nu }{\omega}\,,
\end{align}
where $e_{(k)}^\mu=\dfrac{k^\mu}{\omega}-u^\mu$ represents a unit spacelike vector and $\omega=-u_\mu k^\mu$. Under the action of the projection operator, the components of the polarization tensor within the subspace $\tilde{\mathcal{S}}$ can be expressed as follows
\begin{align}\label{subs}
    \tilde{\mathcal{S}}^{ij}= \dfrac{1}{\nu^3}
    \begin{pmatrix}
        {I_\nu}+{Q_\nu}
        & {U_\nu}+i{V_\nu} \\
        {U_\nu}-i{V_\nu}
        &{I_\nu}-{Q_\nu}
    \end{pmatrix}
    =\begin{pmatrix}
        \mathcal{I}+\mathcal{Q}&\mathcal{U}+i\mathcal{V}\\
        \mathcal{U}-i\mathcal{V}&\mathcal{I}-\mathcal{Q}\,,
    \end{pmatrix}
\end{align}
where we have introduced the quantity \(\mathcal{I} = \dfrac{I_\nu}{\nu^3}\), along with similar quantities \(\mathcal{Q}\), \(\mathcal{U}\), and \(\mathcal{V}\).

\section{Emission, absorption, and Faraday rotation coefficient \label{sec:FC}}
In this study, we examine polarized synchrotron emission, utilizing coefficients derived from the thermal distribution of electrons. The fitting formula we employ is sourced from \cite{Dexter:2016cdk}. The specific results we utilize are consistent with those in \cite{Bronzwaer:2020kle}, and all expressions are presented in CGS units.

The emission coefficients are given as follows:
\begin{align}
j_I &= \dfrac{n_e e^2 \nu}{2\sqrt{3}c \theta_e^2} I_I(x)\,, \\
j_Q &= \dfrac{n_e e^2 \nu}{2\sqrt{3}c \theta_e^2} I_Q(x)\,, \\
j_V &= \dfrac{2n_e e^2 \nu}{3\sqrt{3}c \theta_e^3 \tan\theta_B} I_V(x)\,,
\end{align}
where \( n_e \) denotes the number density of electrons, \( e \) stands for the elementary charge, \( c \) represents the speed of light, \( \theta_B \) symbolizes the angle between the wave vector and the magnetic field, and \( \theta_e = \dfrac{k_B T_e}{m_e c^2} \) characterizes the dimensionless electron temperature. Here, \( k_B \) signifies the Boltzmann constant, and \( T_e \) denotes the thermodynamic temperature. The parameter \( x \) denotes the ratio of the photon frequency to $\nu_c$,
\begin{align}
x = \dfrac{\nu}{\nu_c}, \qquad
\nu_c = \dfrac{3eB \sin\theta_B \theta_e^2}{4\pi m_e c}\,.
\end{align}
Here, \( B \) represents the magnitude of the magnetic field within the fluid system. The expressions for \( I_I \), \( I_Q \), and \( I_V \) are provided by the fitting functions as follows:
\begin{align}
    I_I(x)&=2.5651\pa{1+1.92 x^{-1/3}+0.9977x^{-2/3}}\ee^{-1.8899x^{1/3}}\,,\\
    I_Q(x)&=2.5651\pa{1+0.93193 x^{-1/3}+0.499873x^{-2/3}}\ee^{-1.8899x^{1/3}}\,,\\
    I_V(x)&=\pa{1.81348 x^{-1}+3.42319 x^{-2/3}+0.0292545 x^{-1/2}+2.03773 x^{-1/3}  }\ee^{-1.8899x^{1/3}}\,.
\end{align}
In the context of hot electron distribution, the absorption process adheres to Kirchhoff's law. This law asserts that all absorption coefficients must satisfy the following relationship:
\begin{align}
    a_\nu = \frac{j_\nu}{B_\nu}\,,
\end{align}
where \( B_\nu \) represents the Planck blackbody radiation function.

Finally, the Faraday rotation coefficients are given by the following expressions:
\begin{align}
    r_Q &= \dfrac{n_e e^4 B^2 \sin^2 \theta_B}{4\pi^2 m_e^3 c^3 \nu^3} f_m(X) + \left( \dfrac{K_1(\theta_e^{-1})}{K_2(\theta_e^{-1})} + 6\theta_e \right), \\
    r_V &= \dfrac{n_e e^3 B \cos \theta_B}{\pi m_e^2 c^2 \nu^2} \dfrac{K_0(\theta_e^{-1}) - \Delta J_5(X)}{K_2(\theta_e^{-1})}\,,
\end{align}
where \( K_n(x) \) is the modified Bessel function of the second kind of order \( n \), and
\begin{align}
    X = \left( \dfrac{3}{2\sqrt{2}} \times 10^{-3} \dfrac{\nu}{\nu_c} \right)^{-1/2}.
\end{align}
The remaining functions \( f_m \) and \( \Delta J_5 \) are fitting functions, given by:
\begin{align}
    f_m(X) &= 2.011 \exp\left( -\dfrac{X^{1.035}}{4.7} \right) - \cos\left( \dfrac{X}{2} \right) \exp\left( -\dfrac{X^{1.2}}{2.73} \right) - 0.011 \exp\left( -\dfrac{X}{47.2} \right) \nonumber \\
    &\quad + \dfrac{1}{2} \left[ 0.011 \exp\left( -\dfrac{X}{47.2} \right) - 2^{-1/3} 3^{-23/6} \pi \times 10^4 X^{-8/3} \left( 1 + \tanh\left( 10 \ln \dfrac{X}{120} \right) \right) \right]\,,
\end{align}
and
\begin{align}
    \Delta J_5(X) = 0.4379 \ln\left( 1 + 0.001858 X^{1.503} \right)\,.
\end{align}

\section{Polarization using the PW constant}\label{BB}

In a Type D spacetime, the PW constant can be utilized for the computation of linear polarization. Within Kerr spacetime, there exists a conserved quantity along photon geodesics, represented by the equation \cite{Lee:2022rtg}:
\begin{align}
\kappa = \kappa_1 + i\kappa_2 = 2k^\mu f^\nu \left( \hat{l}_{[\mu}\hat{n}_{\nu]} - \hat{m}_{[\mu}\hat{\bar{m}}_{\nu]} \right) \left( r - ia\cos\theta \right)\,, \label{eqn:PWconstant}
\end{align}
where, \( k^\mu \) indicates the 4-wave vector of the light ray, \( f^\mu \) denotes the unit vector along the polarization direction, and \(\{\hat{l}, \hat{n}, \hat{m}, \hat{\bar{m}}\}\) are the Newman-Penrose tetrads given by:
\begin{align}
\hat{l}^\mu &= \dfrac{1}{\sqrt{2\Delta \Sigma}}\left((r^2 + a^2)\partial_t + \Delta\partial_r + a\partial_\phi\right) \nonumber \\
\hat{n}^\mu &= \dfrac{1}{\sqrt{2\Delta \Sigma}}\left((r^2 + a^2)\partial_t - \Delta\partial_r + a\partial_\phi\right) \nonumber \\
\hat{m}^\mu &= \dfrac{1}{\sqrt{2}\left(r + ia\cos\theta\right)} \left(ia\sin\theta\partial_t + \partial_\theta + \dfrac{i}{\sin\theta}\partial_\phi \right)\,.
\end{align}
The polarization vector in the coordinate system of the fluid is given by:
\begin{align}
f^{\mu} = f^{(a)}e_{(a)}^\mu \,, \qquad f^{(1)} = 1 \,, \qquad f^{(0)} = f^{(2)} = f^{(3)} = 0\,.
\end{align}
At the observer's location, by utilizing Equation \ref{eqn:PWconstant} and the two gauge conditions \( u_\mu f^\mu = k_\mu f^\mu = 0 \), one can determine the polarization direction \( f^\mu \) at the observer. Additionally, by utilizing the conserved quantity \( \mathcal{S} = S_\nu/\nu^3 \), one can compute the linearly polarized intensity \( \mathcal{I}_{\text{pol}} \) on the observer's screen along with the total intensity \( \mathcal{I} \):
\begin{align}
\mathcal{Q} &= \mathcal{I}_{\text{pol}}\left[(f^{(1)})^2 - (f^{(2)})^2\right] \,, \quad
\mathcal{U} = 2\mathcal{I}_{\text{pol}}f^{(1)}f^{(2)}\,.
\end{align}

\bibliographystyle{utphys}
\bibliography{reference}

\end{document}